\documentclass[submission, Proceedings]{SciPost}

\begin{document}

\begin{center}{\Large \textbf{
Measurement of the spectral function for the 
$\tau^-\to K^-K_S\nu_{\tau}$ decay in BABAR experiment  }}
\end{center}

\begin{center}
S. I. Serednyakov \textsuperscript{1*} \\
on behalf of the BABAR collaboration
\end{center}

\begin{center}
{\bf 1} Novosibirsk State University \\
  Budker Institute of Nuclear Physics \\
Novosibirsk 630090 Russia
\\
* seredn@inp.nsk.su
\end{center}

\begin{center}
\today
\end{center}

\definecolor{palegray}{gray}{0.95}
\begin{center}
\colorbox{palegray}{
\begin{tabular}{rr}
\begin{minipage}{0.05\textwidth}
\includegraphics[width=8mm]{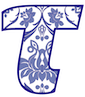}
\end{minipage}
 &
 \begin{minipage}{0.82\textwidth}
 \begin{center}
 {\it Proceedings for the 15th International Workshop on
 Tau Lepton Physics,}\\
 {\it Amsterdam, The Netherlands, 24-28 September 2018} \\
 \href{https://scipost.org/SciPostPhysProc.1}{\small
 \sf scipost.org/SciPostPhysProc.Tau2018}\\
 \end{center}
 \end{minipage}
 \end{tabular}
  }
\end{center}


\section*{Abstract}
{\bf The  decay $\mathbf{\tau^{-}\to K^{-}K_S\nu_{\tau}}$ has been studied 
using $\mathbf{430\times10^6 ~ e^+e^-\to \tau^+\tau^-}$
events produced at a center-of-mass energy around 10.6 GeV
 at the PEP-II collider and studied with the BABAR detector.
The mass spectrum of the $\mathbf{K^{-}K_S}$ system has been measured and the 
spectral function has been obtained. The 
measured branching fraction 
$\mathbf{{\cal B}(\tau^{-}\to K^{-}K_S\nu_{\tau})=
(0.739\pm 0.011(\rm stat.)\pm 0.020(\rm syst.))}$ $\mathbf{\times 10^{-3}}$
is found to be in agreement with earlier measurements.
}



\section{Introduction}
\label{sec:intro}
The $\tau$ lepton provides a remarkable laboratory 
for studying many open questions in particle physics. 
With a large statistics of  about $10^9$ $\tau_s$ produced in $e^+e^-$ 
annihilation at the BABAR experiment, 
various aspects can be  studied,
for example, improving  the precision of 
spectral functions describing 
the mass distribution of  the hadronic decays of the  $\tau$.
In this work, we analyze 
the $\tau^{-}\to K^{-}K_S\nu_{\tau}$ decay\footnote{Throughout this paper,
inclusion of charge-conjugated channels  is implied.} 
and measure the spectral function of this channel defined as ~\cite{CVC}
\begin{equation}
V(q)=\frac{m_{\tau}^8}{12\pi C(q) |V_{ud}|^2}
\frac{{\cal B}(\tau^- \to K^-K_S\nu_\tau)}
{{\cal B}(\tau^- \to e^-\bar{\nu_e}\nu_\tau)}
\frac{1}{N}\frac{dN}{dq},
\label{eq1}
\end{equation}
where $m_{\tau}$ is the $\tau$ mass \cite{PDG},
$q\equiv m_{K^-K_S}$ is the invariant mass of the $K^-K_S$
system,  $V_{ud}$ is an 
element of the CKM (Cabibbo-Kobayashi-Maskava) matrix \cite{PDG}, 
$(dN/dq)/N$ is the normalized $K^-K_S$ mass 
spectrum,  and $C(q)$ is 
the phase space correction factor given by the following formula:
\begin{equation}
C(q)=q(m_{\tau}^2-q^2)^2(m_{\tau}^2+2q^2).
\label{eq2}
\end{equation}

\begin{figure}[t]
\includegraphics[width=0.47\textwidth]{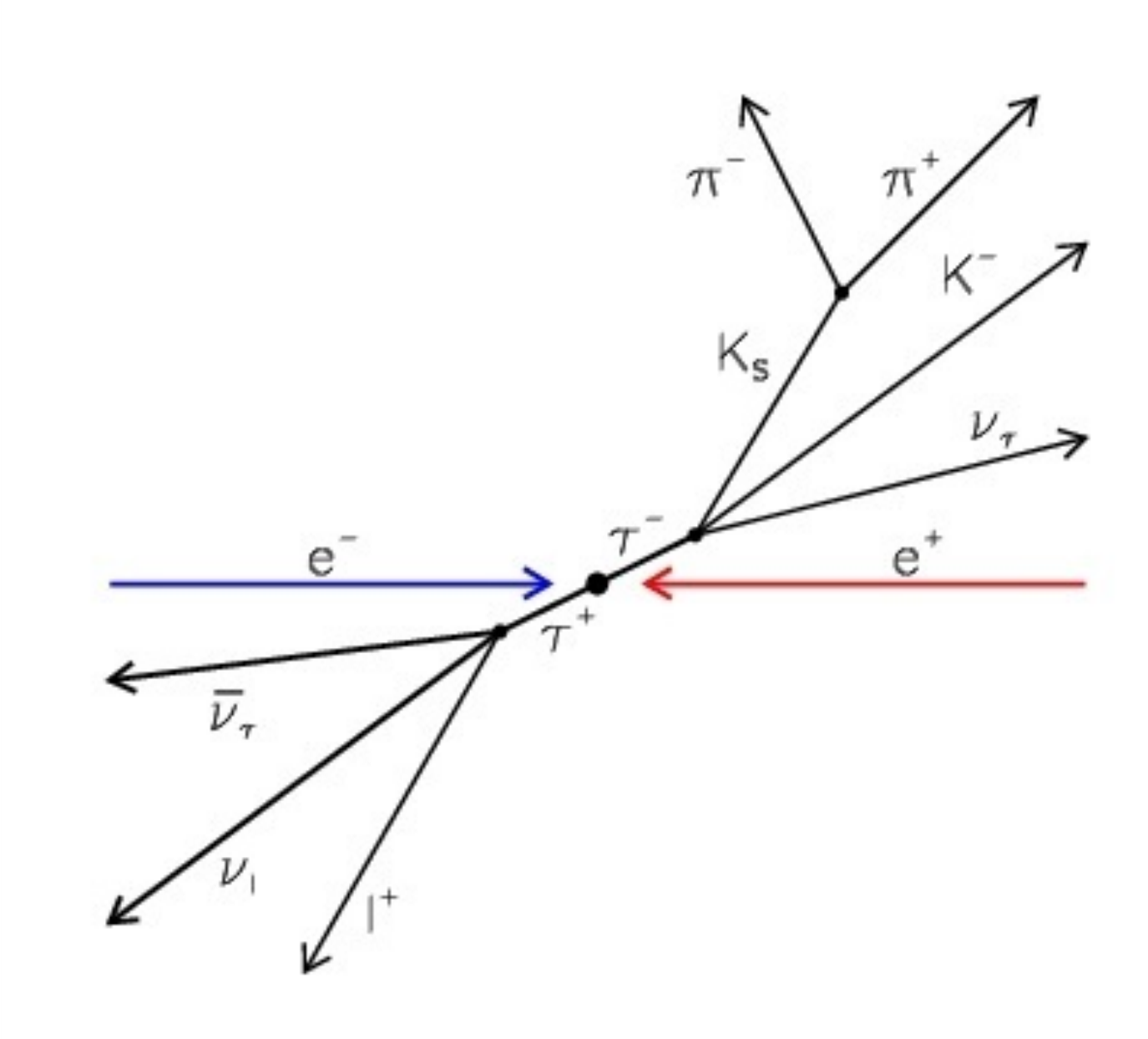}
\hfill
\includegraphics[width=0.47\textwidth]{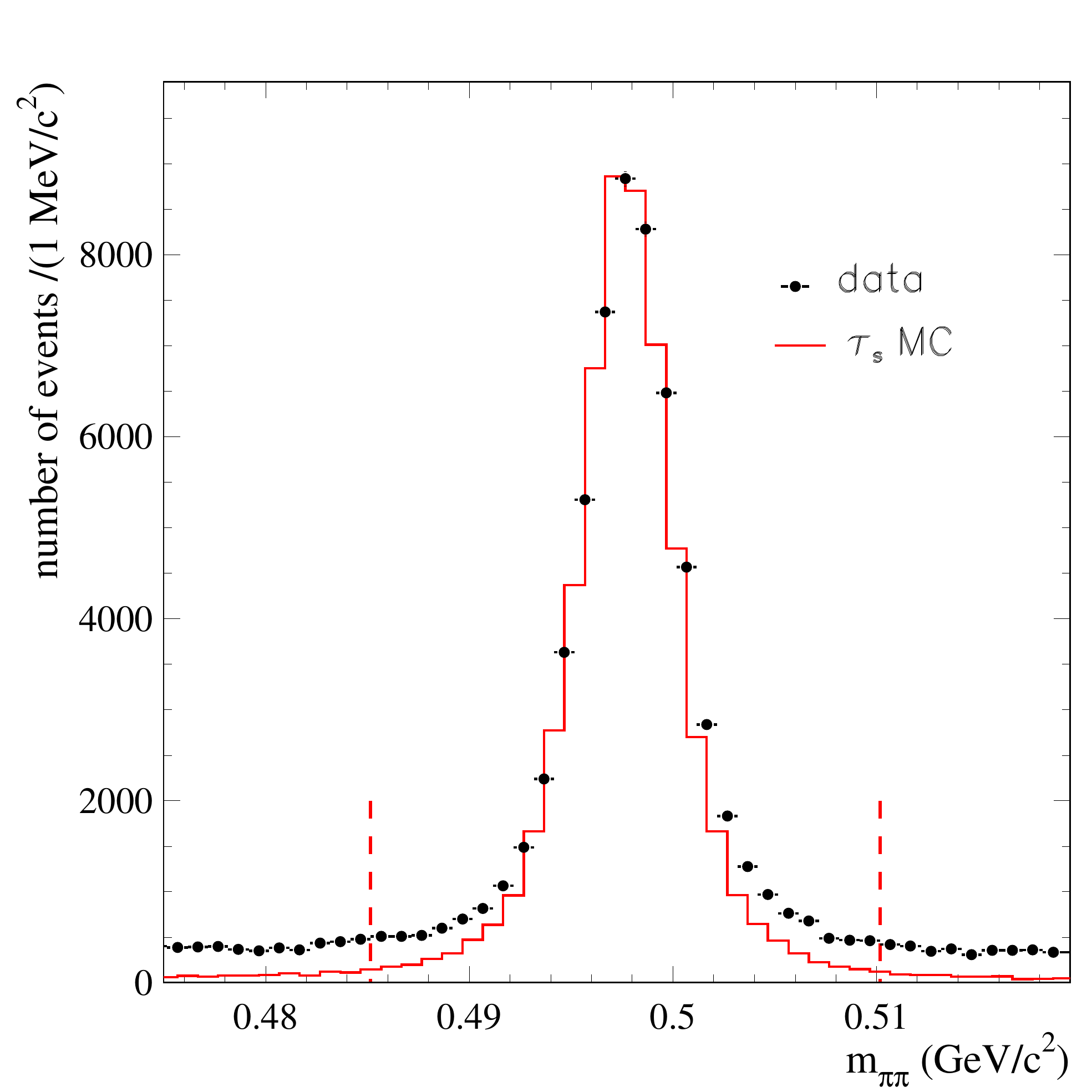}
\parbox[t]{0.47\textwidth} {
\vspace*{-7mm}
\caption {Schematic view of the $\tau$ decay chains in
$e^+e^-\to\tau^+\tau^-$ events selected for this
analysis. Lepton $l^+$ can be electron or muon.} 
\label{decays} }
\hfill
\parbox[t]{0.47\textwidth} {
\vspace*{-7mm}
\caption {The $\pi^+\pi^-$ mass spectrum for $K_S$
candidates in data (points with errors) and signal
simulation (histogram).
Between the two vertical lines there is a signal region
used in the procedure of non-$K_S$ background subtraction. }
\label{ksmas} }
\end{figure}

The branching fraction for the $\tau^{-}\to K^{-}K_S\nu_{\tau}$ decay has been
measured with relatively high (3\%) precision by the Belle 
experiment~\cite{Belle}. The $K^-K_S$ mass spectrum 
was measured by the CLEO experiment~\cite{CLEOt}. In the CLEO analysis, 
a data set of $2.7\times10^6$ produced $\tau$ pairs was used, and 
about 100  events in the decay channel $\tau^-\to K^-K_S\nu_{\tau}$
were selected.
In this work~\cite{PresArt}, using  about $\sim$ 10$^9$   $\tau$ leptons, 
we significantly improve upon the measurement of the spectral function for 
the $\tau^-\to K^-K_S\nu_{\tau}$ decay.

\section{Data used in the analysis}
\label{sec:Data}
We analyze a data sample corresponding to an integrated
luminosity of 468 fb$^{-1}$ recorded with the BABAR
detector~\cite{BABAR1},\cite{BABAR2} at the SLAC PEP-II 
asymmetric-energy $e^+e^-$ collider. In the laboratory frame, 
the energy of electron and positron beams is 9 and 3.1 GeV, respectively.

For simulation of $e^+e^-\to \tau^+\tau^-$ events 
the KK2f Monte Carlo generator~\cite{kk2f} is used, which includes
higher-order radiative corrections to the Born-level process.
Decays of $\tau$ leptons are simulated using  the 
Tauola package~\cite{Tauola}.
Two separate samples of simulated $e^+e^-\to \tau^+\tau^-$ events are used: 
a generic sample with $\tau$ decaying to all significant final states, 
and the signal channel where 
$\tau^+\to l^+\nu_l\bar{\nu}_\tau$, $l=e$ or $\mu$ and 
$\tau^-\to K^-K_S\nu_\tau$.
To estimate backgrounds, we use a sample of simulated generic  
$e^+e^-\to \tau^+\tau^-$ events after excluding the signal decay channel 
($\tau^+\tau^-$ background) and a sample containing all events arising from 
$e^+e^-\to q\bar{q}$, $q=u,d,s,c$ and $e^+e^-\to B\bar{B}$ processes 
($q\bar{q}$ background).
The $q\bar{q}$ background  events with $q=u,d,s,c$
are generated using the JETSET generator~\cite{JETSET}, while $B\bar{B}$ 
events are simulated with EVTGEN~\cite{EVTGEN}. The detector response is 
simulated with GEANT4 ~\cite{GEANT4}. The equivalent luminosity of the 
simulated sample is 2-3 times higher than the  integrated luminosity in data.

\section{Event selection}
\label{sec:Sel}
We select $e^+e^-\to \tau^+\tau^-$ events with  the $\tau^+$  decaying
leptonically ($\tau^+\to l^+\nu_l\bar{\nu}_\tau$, $l=e$ or $\mu$) and the
$\tau^-$ decaying to $K^-K_S\nu_\tau$.
Such events referred to as signal events below.
The $K_S$ candidate is detected in the
$K_S\to\pi^+\pi^-$ decay mode. The topology of events to be selected is
shown in Fig.~\ref{decays}. Unless otherwise stated, all
quantities are measured in the laboratory frame.
The selected events must satisfy the following requirements:
\begin{itemize}
\renewcommand{\labelitemi}{--}
\item The total number of charged tracks, $N_{\rm trk}$,  must be four 
and the total charge of the event must be zero.

\item  Among the four charged tracks there must be an identified lepton   
(electron or muon) and an identified kaon of opposite charge. 

\item To reject non $\tau^+\tau^-$ signal backgrounds,
the lepton candidate must have a momentum above 1.2 GeV/$c$, the momentum in
the center-of-mass frame (c.m. momentum) must be smaller than 4.5 GeV/$c$,
and the cosine of the lepton polar angle $|\cos{\theta_{l}}|$  
must be below 0.9. 

\item  To suppress background from charged pions,
the charged kaon candidate must have a momentum, $p_{K}$,  above 0.4 GeV/$c$
and below 5 GeV/$c$, and the cosine of its polar angle must lie between -0.7374
and 0.9005.

\item The two remaining tracks, assumed to be pions,
form the $K_S$ candidate. The $\pi^+\pi^-$ invariant mass must lie 
within 25 MeV/$c^2$ of the nominal $K_S$ mass, 497.6 MeV/$c^2$. The 
$K_S$ flight length $r_{K_S}$, measured as the distance between
the $\pi^+\pi^-$ vertex and the collision point, must be larger than 1 cm.

\item  The total energy in neutral clusters, $\Sigma E_{\gamma}$, must be less
than 2 GeV. 
Here, a neutral cluster is defined as a local energy deposit
in the calorimeter with energy above 20 MeV and no associated charged
track.

\item The magnitude of the thrust \cite{thrust1,thrust2}
for the event,  calculated using 
charged tracks only,   must be greater than 0.875.

\item  The angle between the momentum of the lepton and the direction
of the hadronic final state in the c.m.\ frame  should be between 110 
and 180 degrees.

\end{itemize} 
  The chosen selection requirements are close to those used in previous
$\tau$ studies in BABAR~ \cite{taucut}.
As a result of applying these cuts
the $\tau$ background is suppressed by 3.5 orders of magnitude,
and the $q\bar{q}$ background by 5.5 orders.

\section{Detection efficiency}
\label{sec:deteff}
The detection efficiency obtained after applying the selection criteria is
calculated using signal Monte Carlo simulation as a function of the 
true $m_{\it K^-K_S}$
mass.   The efficiency is 
weakly dependent on $m_{K^-K_S}$. The average 
efficiency over the mass spectrum is about 13\%. 
It should be noted that the  $K^-K_S$ mass resolution is 
about 2-3 MeV/$c^2$,  significantly smaller than the size of the mass bin
(40 MeV/$c^2$) used in our analysis.  
Therefore, in the following we neglect the effects of the finite 
$K^-K_S$   mass resolution.

To correct for the imperfect simulation of the kaon identification
requirement, 
the particle identification PID efficiences have been compared for
data and simulation on high purity control samples of kaons from 
$D^{\star +}\to\pi^+D^0,~D^0\to K^-\pi^+$ decays \cite{BBPhys}.
We correct
the simulated efficiency using the measured ratios of the
efficiencies measured in data and Monte Carlo, in bins of the kaon
candidate momentum and polar angle. 
The resulting correction factor is small $\sim$ 1\% and weakly
depends on $m_{K^-K_S}$.

\section{Subtraction of non-$K_S$ background}
\label{sec:specks}
The  $\pi^+\pi^-$ mass spectra for $K_S$ candidates in data 
and simulated signal events are shown in Fig.~\ref{ksmas}.
The data spectrum consists of a peak at the $K_S$ mass and a flat 
background. 
To subtract the non-$K_S$ background, the following procedure is used.
The signal region is set to $\pi^+\pi^-$ masses within 0.0125 GeV/$c^2$ of the
$K_S$ mass (indicated by arrows in  Fig.~\ref{ksmas}), 
and the sidebands are set to between 
0.0125 and 0.0250 GeV/$c^2$ away from the nominal
$K_S$ mass. Let $\beta$ be the
fraction of events with a true $K_S$ that fall in the
sidebands, and let $\alpha$ be the fraction of  non-$K_S$ events
that fall in the sidebands. The total number of events in the signal
region plus the sidebands, $N$, and the number of events in the
sidebands, $N_{sb}$, depend on  the number 
of true $K_S$, $N_{K_S}$,
and the number of non-$K_S$ background events, $N_{b}$
according to the following relation :

\begin{subequations}
\begin{equation} N = N_{K_S} + N_{b},
\end{equation}
\begin{gather} N_{sb} = \alpha\cdot N_{b} + \beta\cdot N_{K_S}
\end{gather}
\end{subequations}

Therefore:
\begin{equation}
N_{K_S}=(\alpha N - N_{sb})/(\alpha-\beta).
\label{eq6}
\end{equation}
The value of $\beta$ is determined  using $\tau$  signal simulation. 
It is found to be nearly  independent of the $m_{K^-K_S}$ 
mass and is equal to   0.0315~$\pm$~0.0015.
The value of  $\alpha$ is expected to be 0.5 for a uniformly distributed 
background.  This  is   
consistent with the value 0.499~$\pm$~0.005 obtained on simulated $\tau^+\tau^-$
background events. 
The non-$K_S$ background is subtracted in each $m_{K^-K_S}$ bin. Its
fraction is found to be about 10\% of the selected events with  
$m_{K^-K_S}$ near and below 1.3 GeV/$c^2$ and increases up to 50\% above
1.6 GeV/$c^2$. 

\section{Subtraction of $\tau$-background with a $\pi^0$}
\label{sec:rmspe}
Although the studied process $\tau^-\to K^-K_S\nu_{\tau}$
is not supposed to contain a $\pi^0$ in the final state,   
some events from background processes with a $\pi^0$ pass the selection 
criteria. In the following, we describe how the $\pi^0$ background 
contribution is subtracted.

According to the simulation, the number of signal and
$\tau$-background events are of the same order of magnitude. 
The $\tau^+\tau^-$ background 
consists of events with the decay $\tau^-\to K^-K_S\pi^0\nu_{\tau}$ 
(79\%), events with a misidentified kaon from decays    
$\tau^-\to \pi^-K_S\nu_{\tau}$ (10\%) and 
$\tau^-\to \pi^-K_S\pi^0 \nu_{\tau}$ (3\%), 
and events with a misidentified lepton mainly from the decays
$\tau^+\to \pi^+\bar{\nu}_{\tau}$ and
$\tau^+\to \pi^+\pi^0\bar{\nu}_{\tau}$ (7\%).
Thus, more than 80\% of the background events contain a $\pi^0$ 
in the final state. The hadronic mass spectra for $\tau$ decays with
a $\pi^0$ are not well known, so we use the experimental data to 
subtract this background.

   The $\tau$ background without a $\pi^0$ 
($\tau^-\to \pi^-K_S\nu_{\tau}$, $\tau^+\to \pi^+\bar{\nu}_{\tau}$) and 
$q\bar{q}$ background are simulating well. Therefore, this background 
is   subtracted using Monte Carlo simulation.

To subtract the $\pi^0$ background, the selected events are
divided into two classes, without and with a  
$\pi^0$ candidate, which is
defined as a pair of photons with an invariant mass in the range 
$100-160$ MeV/$c^2$. 

\begin{figure}[t]
\includegraphics[width=0.47\textwidth]{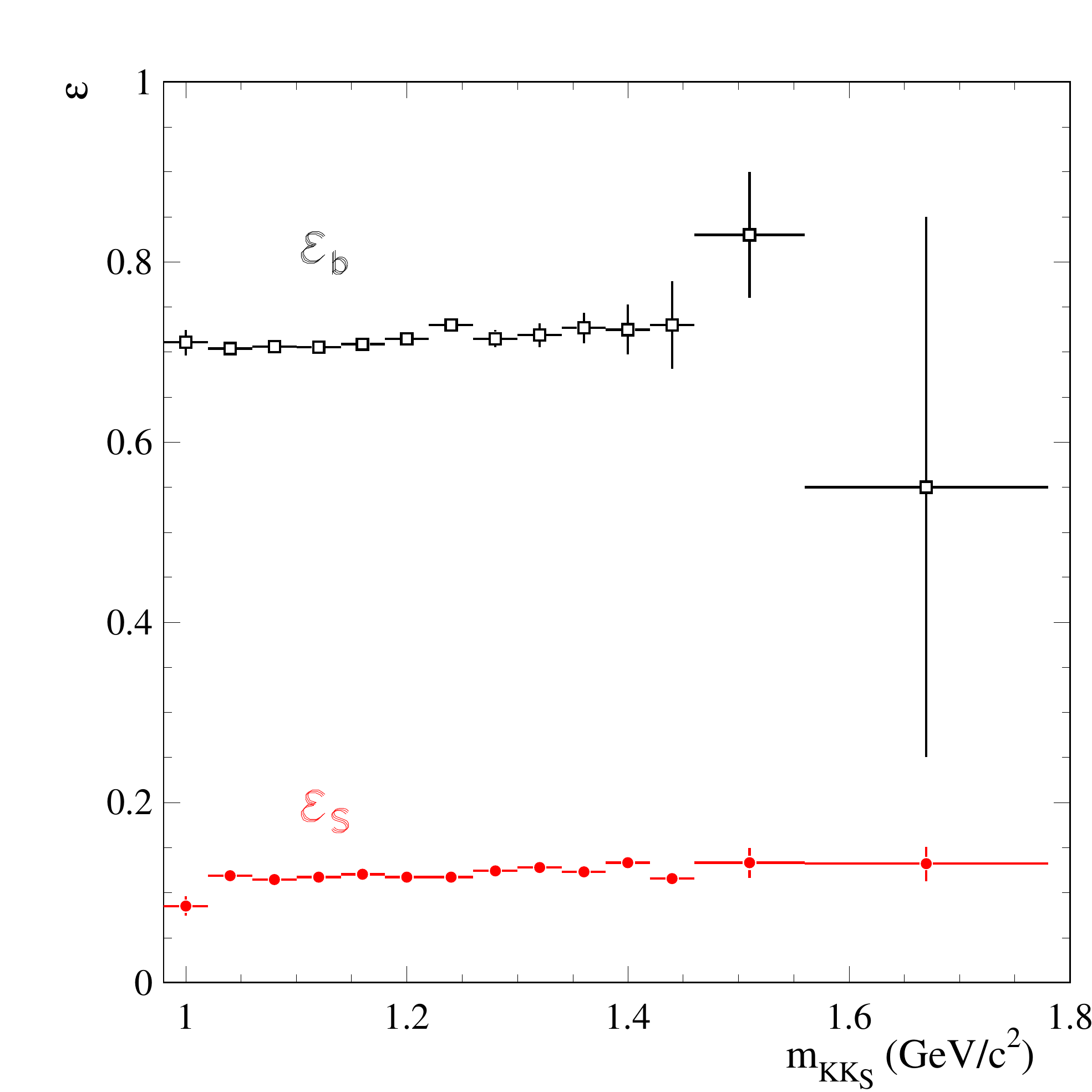}
\hfill
\includegraphics[width=0.47\textwidth]{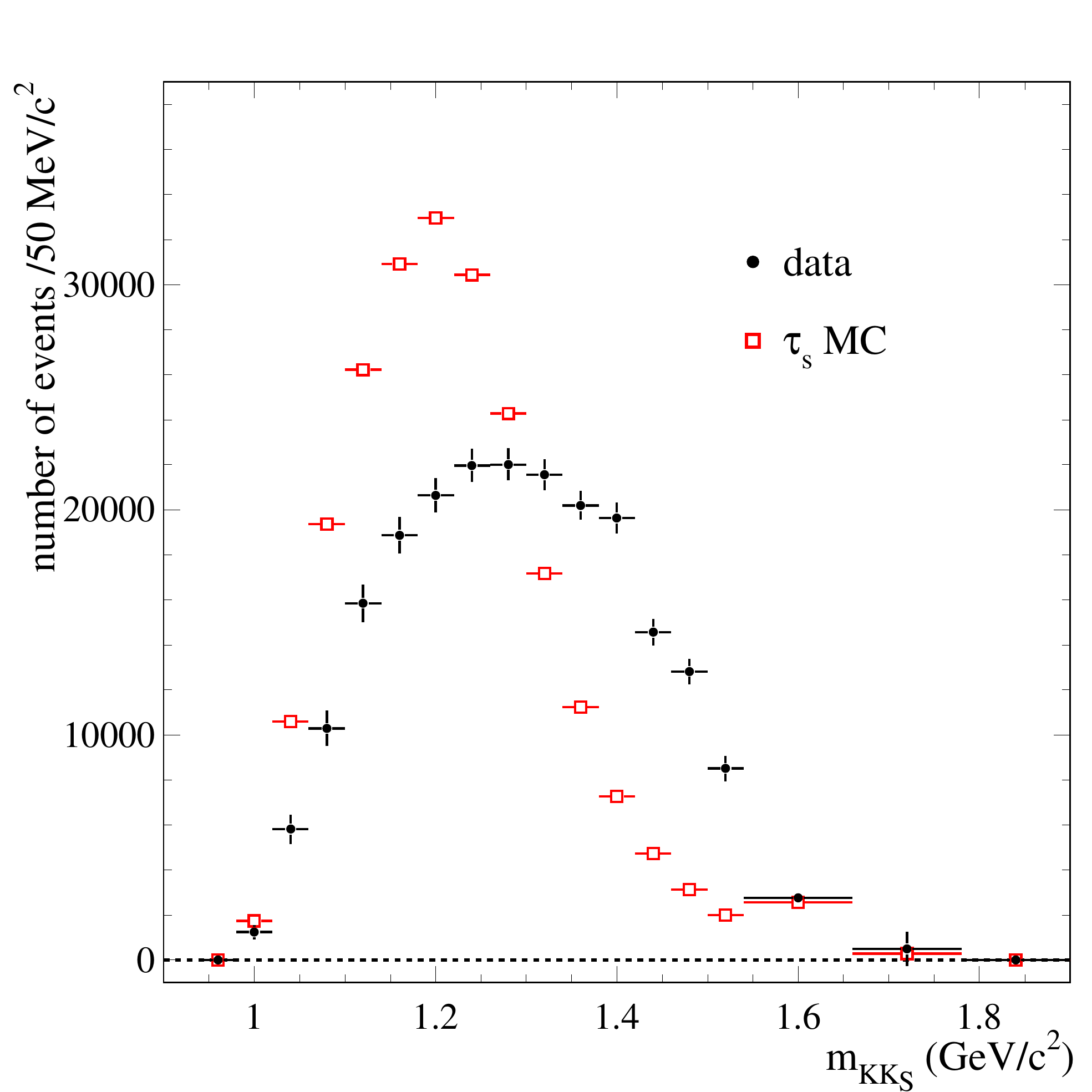}
\parbox[t]{0.47\textwidth} {
\vspace*{-5mm}
\caption { The probabilities $\epsilon_{ s}$ and
$\epsilon_{ b}$ used in Eqs.~(\ref{eq8}a, \ref{eq8}b) as functions of the
$K^-K_S$ mass, measured on simulated events.}
\label{efcut} }
\hfill
\parbox[t]{0.47\textwidth} {
\vspace*{-5mm}
\caption { Measured $m_{ K^-K_S}$  spectra for signal events  in comparison
with the Monte Carlo  simulation.}
\label{dtmc} }
\end{figure}

On the resulting sample, the numbers of signal ($N_{s}$) and background
$\tau^+\tau^-$ events containing a $\pi^0$ candidate ($N_{ b}$) are obtained
in each $m_{ K^-K_S}$ bin:
\begin{subequations}
\begin{equation}
N_{0\pi^0}=(1-\epsilon_{s})N_{s}+(1-\epsilon_{b})N_{b},
\end{equation}
\begin{gather}
N_{1\pi^0}=\epsilon_{s}N_{s}+\epsilon_{b}N_{b},
\end{gather}
\label{eq8}
\end{subequations}
where $N_{0\pi^0}$ and $N_{1\pi^0}$ are the numbers of selected data events
with zero and at least one $\pi^0$ candidate, and 
$\epsilon_{ s}$ ($\epsilon_{ b}$) is the probability for signal  
(background) $\tau^+\tau^-$ events to be found in events 
with at least one $\pi^0$ candidate calculated using Monte Carlo simulation. 
The values  $\epsilon_{ s}$ and $\epsilon_{ b}$ for each bin 
in  $m_{ K^-K_S}$ are measured in Monte
Carlo by counting how many signal and background event candidates
contain a $\pi^0$ candidate.
Figure \ref{efcut} shows the $\epsilon_{ s}$ and $\epsilon_{ b}$ measured 
in Monte Carlo as a function of $m_{ K^-K_S}$.
These efficiencies are corrected to take into account 
the difference between data and Monte Carlo.

With  these corrected values for
$\epsilon_{ s}$ and $\epsilon_{ b}$ we solve 
Eqs.~(\ref{eq8}a, \ref{eq8}b) for each
$K^-K_S$ mass bin and obtain mass spectra for signal ($N_{ s}$)
and background ($N_{ b}$). The efficiency corrected signal mass 
spectrum  is shown in Fig.~\ref{dtmc} in comparison with the simulation.
We find a substantial difference between data and simulation for
the signal spectrum. The result is not affected by inaccuracies 
of the simulation since it doesn't depend on the normalization 
of the simulated $m_{ K^-K_ S}$  spectrum.

\section{ Systematic uncertainties}

  The uncertainty from non-$K_S$ background subtraction (0.4\%) 
is estimated by varying the coefficients of $\alpha$ and $\beta$
within their uncertainties. This uncertainty is 
independent on the $K^-K_S$ mass.
The PID correction
uncertainty due to data-Monte Carlo simulation difference
in particle identification
is taken to be 0.5\%, independent of the $K^-K_S$ mass.
The uncertainty on how well the Monte Carlo simulates the tracking
efficiency is estimated to be 1\%.
We take the observed difference between data and Monte Carlo 
near the end point $M_{ K^-K_S}=m_\tau$ 
as an uncertainty on the $q\bar{q}$ background. 
This leads to an uncertainty 
on ${\cal B}(\tau^-\to K^-K_S\nu_{\tau})$ of 0.5\%. 
The uncertainty associated with the subtraction 
of the $\tau^+\tau^-$ background
with  $\pi^0_s$ is estimated to be 2.3\%.

The systematic uncertainties from different sources
are combined in quadrature. The total systematic uncertainty for
the branching fraction ${\cal B}(\tau^-\to K^-K_S\nu_{\tau})$ is 2.7\%.
The systematic uncertainties for the mass spectrum are listed
in Table~\ref{tab-2}. They gradually decrease from $\simeq$9\% at 
$m_{ K^-K_S}=1$ GeV/$c^2$ to 1.5\% at $m_{ K^-K_S}=m_\tau$.
Near the maximum of the mass spectrum (1.3 GeV/$c^2$)
the uncertainty is about 2.5\%. 

\begin{figure}[t]
\includegraphics[width=0.47\textwidth]{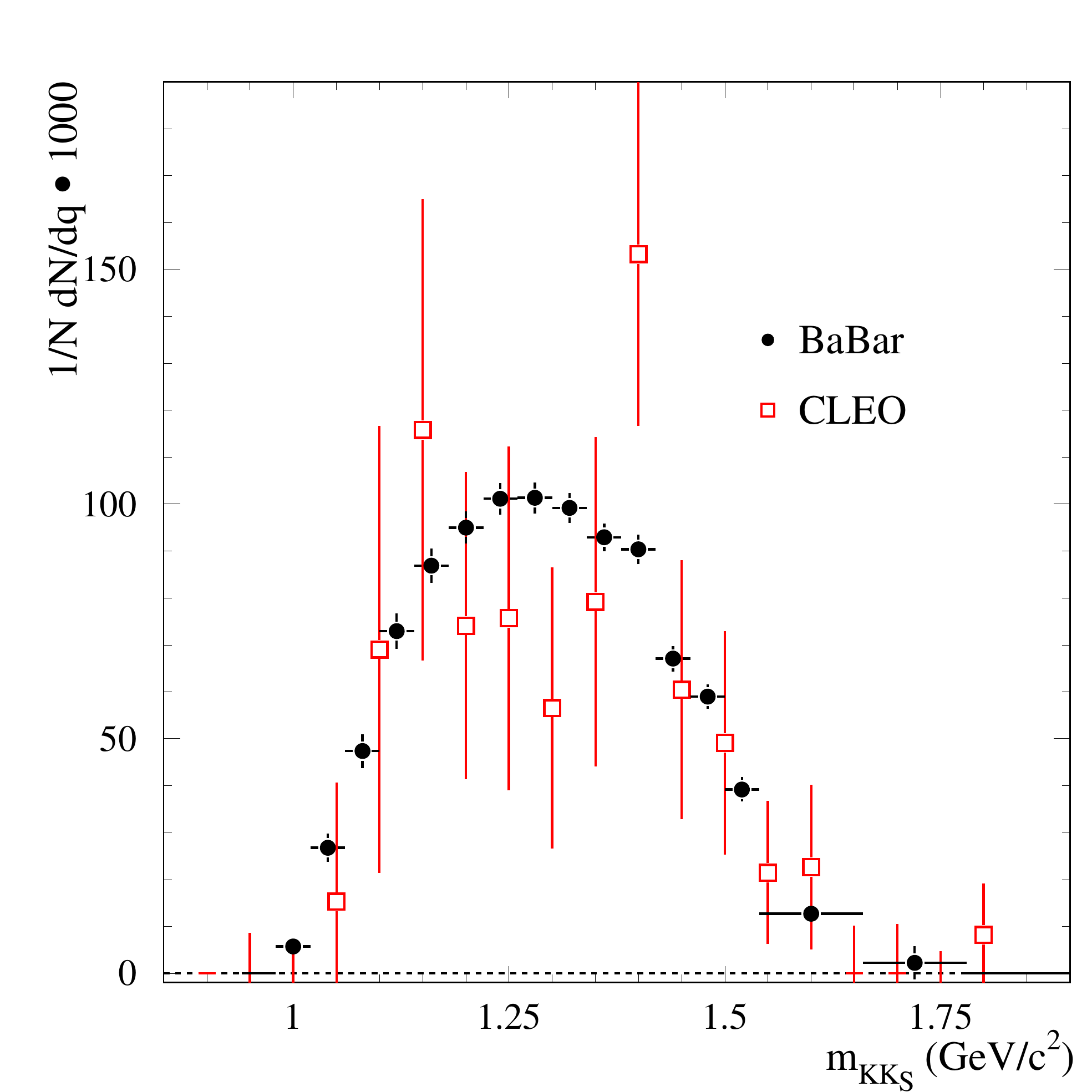}
\hfill
\includegraphics[width=0.47\textwidth]{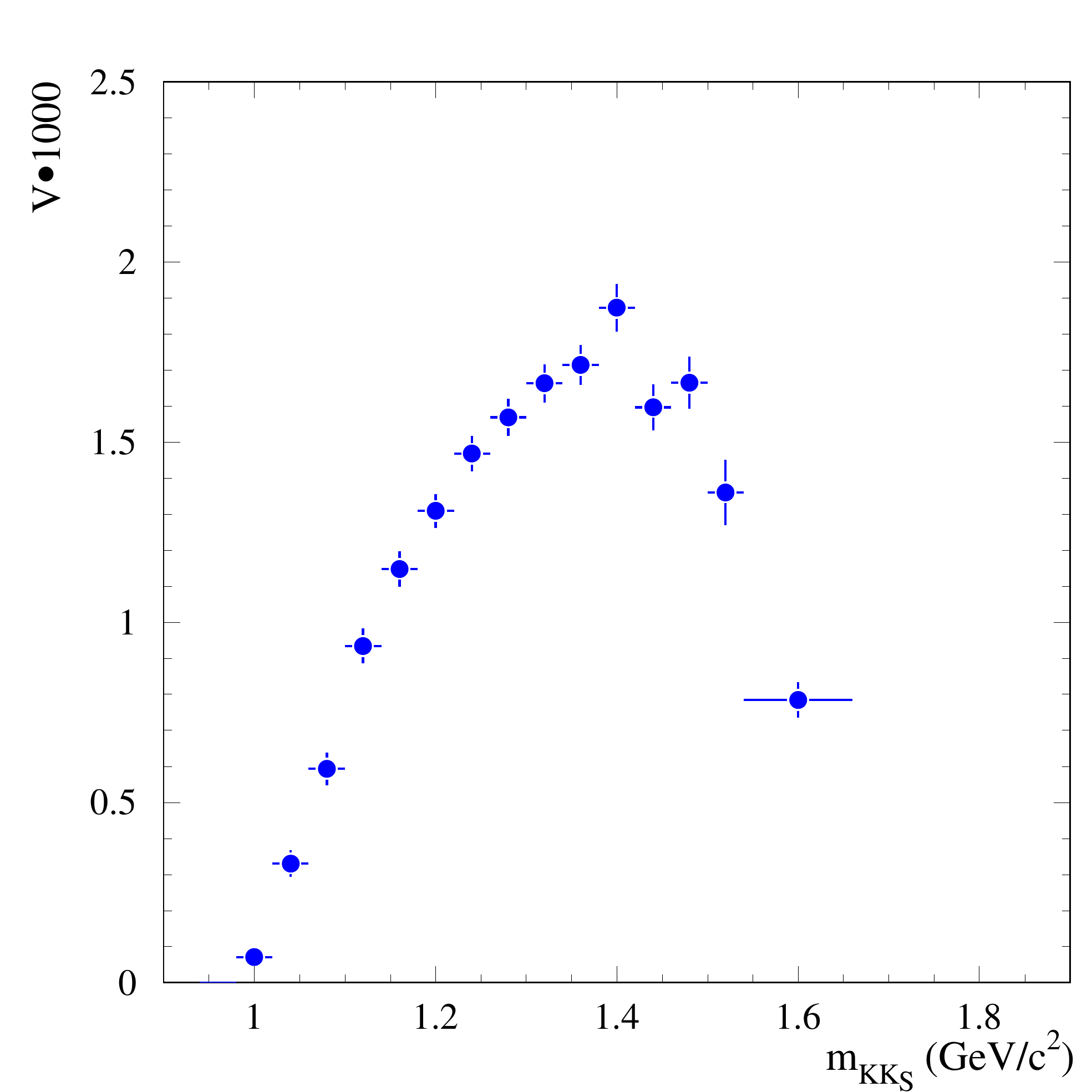}
\parbox[t]{0.47\textwidth} {
\vspace*{-7mm}
\caption { Normalized $K^-K_S$ invariant mass
spectrum for the $\tau^-\to K^-K_S\nu_{\tau}$ decay measured in this work
(filled circles) compared to the CLEO measurement~\cite{CLEOt} (empty squares).
Only statistical uncertainties are shown. }
\label{mspec} }
\hfill
\parbox[t]{0.47\textwidth} {
\vspace*{-7mm}
\caption { Measured spectral function for the
$\tau^- \to K^-K_S\nu_\tau$ decay. Only statistical uncertainties 
are shown.}
\label{vsfcs} }
\end{figure}

\begin{table}
\parbox[h]{0.7\textwidth}{\caption{
Measured spectral function (V) of the $\tau^- \to K^-K_S\nu_{\tau}$
decay,
in bins of $m_{ K^-K_S}$.
The columns report: the range of the bins, the normalized number of
events, the value of the spectral function. The first error is statistical, 
the second systematic.} 
\label{tab-2}}
\begin{tabular}{ccc} \\ \hline
$m_{ K^-K_S}$(GeV/c$^2$)  &$N_s/N_{tot}\times 10^3$ &$V\times10^3$     \\ \hline
$0.98-1.02$  &$5.6 \pm1.4 $ &$0.071\pm0.018\pm0.006 $   \\  
$1.02-1.06$  &$26.0\pm2.7 $ &$0.331\pm0.034\pm0.026 $  \\ 
$1.06-1.10$  &$46.0\pm3.2 $ &$0.593\pm0.042\pm0.042 $   \\  
$1.10-1.14$  &$70.8\pm3.5 $ &$0.934\pm0.046\pm0.056 $  \\ 
$1.14-1.18$  &$84.4\pm3.4 $ &$1.148\pm0.047\pm0.057 $  \\  
$1.18-1.22$  &$92.3\pm3.3 $ &$1.309\pm0.046\pm0.052 $  \\ 
$1.22-1.26$  &$98.2\pm3.2 $ &$1.468\pm0.048\pm0.044 $ \\  
$1.26-1.30$  &$98.4\pm3.2 $ &$1.569\pm0.050\pm0.042 $  \\ 
$1.30-1.34$  &$96.3\pm3.0 $ &$1.663\pm0.052\pm0.042 $  \\  
$1.34-1.38$  &$90.2\pm2.9 $ &$1.715\pm0.052\pm0.039 $   \\ 
$1.38-1.42$  &$87.8\pm3.1 $ &$1.873\pm0.066\pm0.039 $  \\  
$1.42-1.46$  &$65.1\pm2.6 $ &$1.597\pm0.064\pm0.032 $  \\ 
$1.46-1.50$  &$57.3\pm2.5 $ &$1.666\pm0.073\pm0.032$    \\  
$1.50-1.54$  &$38.1\pm2.5 $ &$1.361\pm0.090\pm0.023 $  \\ 
$1.54-1.66$  &$36.9\pm2.4 $ &$0.785\pm0.049\pm0.013 $   \\  
$1.66-1.78$  &$6.6 \pm10.2$ &$0.986\pm1.520\pm0.014  $   \\ \hline
\end{tabular}
\end{table}

\section{The results}
\label{sec:res}
The branching ratio of the $\tau^-\to K^-K_S\nu_{\tau}$ decay is obtained
using the following expression:
\begin{eqnarray}\label{eq11}
{\cal B}(\tau^{-}\to K^{-}K_S\nu_{\tau})=
\frac{N_{\rm exp}}{2LB_{\rm lep}\sigma_{\tau\tau}}=
\nonumber\\(0.739\pm0.011\pm0.020)\times10^{-3},
\end{eqnarray}
where $N_{\rm exp}=223741~\pm~3461$ (error is statistical) 
is the total number of signal events in the
spectrum in Fig.~\ref{mspec}, $L=468.0\pm2.5$ fb$^{-1}$ 
is the BABAR integrated luminosity \cite{rlumi},
$\sigma_{\tau\tau}=0.919\pm0.003$ nb is the 
$e^+e^-\to \tau^+\tau^-$ cross section
at 10.58 GeV ~\cite{kk2f}
and $B_{\rm lep}$=0.3521~$\pm$~0.0006 is the world average 
sum of electronic and muonic branching fractions of the  $\tau$ lepton \cite{PDG}.
The first uncertainty  in (\ref{eq11}) is the statistical, 
the second is systematic.  Our result agrees well with
the Particle Data Group (PDG) value
$(0.740\pm 0.025)\times 10^{-3}$~\cite{PDG}, which is determined mainly
by the recent Belle measurement
$(0.740\pm 0.007\pm 0.027)\times 10^{-3}$~\cite{Belle}.

The measured mass spectrum $m_{ K^-K_S}$ for the $\tau^- \to K^-K_S\nu_{\tau}$ 
decay is shown in Fig.~\ref{mspec} and listed in Table~\ref{tab-2}. 
Our $m_{ K^-K_S}$ spectrum is compared with 
the CLEO measurement~\cite{CLEOt}. The BABAR and
CLEO spectra are in good agreement.
The spectral function $V(q)$ calculated using 
Eq.~(\ref{eq1}) is shown in Fig.~\ref{vsfcs} and listed in Table \ref{tab-2}.
Due to the large error in the mass interval 1.66-1.78 GeV/c$^2$, which exceeds 
the scale of Fig.~\ref{vsfcs}, the value of $V(q)$ in this interval is not
shown in Fig.~\ref{vsfcs}.

\section{Conclusions}
The $K^-K_S$ mass spectrum and vector spectral function in the 
$\tau^-\to K^-K_S\nu_{\tau}$ decay have
been  measured by the BABAR experiment. 
The measured $K^-K_S$ mass spectrum is far more precise than 
CLEO measurement \cite{CLEOt} and the branching fraction
$(0.739\pm 0.011\pm 0.020)\times 10^{-3}$ is comparable to 
Belle's measurement \cite{Belle}.

\section*{ACKNOWLEDGMENTS}
  The author of this talk is grateful to V. Druzhinin and A. Lusiani for useful
discussions. This work in part of data analysis is supported by the
Russian Foundation for Basic Researches  (grant No. 16-02-00327).

\nolinenumbers

\end{document}